\documentclass[prl,twocolumn,superscriptaddress,preprintnumbers,amsmath,amssymb]{revtex4}
\usepackage[dvips]{graphicx}
\usepackage{dcolumn}
\usepackage{color}
\usepackage{bm}

\newcommand{\hatH}{\hat{H}}

\begin{document}

\title{Calorimetric measurement of quantum work}

\author{Jukka P. Pekola}
\affiliation{Low Temperature Laboratory (OVLL), Aalto University School of Science, P.O. Box 13500, 00076 Aalto, Finland}
\author{Paolo Solinas}
\affiliation{Low Temperature Laboratory (OVLL), Aalto University School of Science, P.O. Box 13500, 00076 Aalto, Finland}
\affiliation{Department of Applied Physics, Aalto University School of Science,
P.O. Box 11000, 00076 Aalto, Finland}
\author{Alexander Shnirman}
\affiliation{Institut f\"ur Theorie der Kondensierten Materie, Karlsruhe Institute of Technology, 76128 Karlsruhe, Germany}
\author{Dmitri V.\ Averin}
\affiliation{Department of Physics and Astronomy, Stony Brook University, SUNY, Stony Brook, NY 11794-3800,
USA}

\date{\today}

\begin{abstract}
To define the work performed on a driven quantum system in a physically sound way has turned out to be a truly non-trivial task, except in some special cases of limited applicability \cite{bochkov77, kurchan00, tasaki00, mukamel03, yukawa00, chernyak04, allahverdyan05, engel07,talkner07,campisi11,solinas12}. This topic has been in a focus of intense research recently in the attempts to generalize the classical fluctuation relations \cite{bochkov81,jarzynski97,crooks99,crooks00,jarzynski08,ap11} into the quantum regime. Here we propose and demonstrate that a calorimetric measurement gives both a theoretical and experimental tool to test the Jarzynski equality (JE) and other fluctuation relations in a quantum system, and to analyze the distribution of dissipation in them, based on the very principle of conservation of energy. We focus on an experimentally feasible two-level system, a superconducting Cooper pair box  \cite{averin85,bouchiat98, nakamura99} subject to Landau-Zener interband transitions \cite{landau32,zener32,shevchenko10}.
Because of the small heat capacity and weak relaxation to the phonon bath, the calorimetric measurement on the electron gas (resistor) turns out to be a very feasible experimental method.
\end{abstract}

\maketitle

\begin{figure}
    \includegraphics[width=8.5cm]{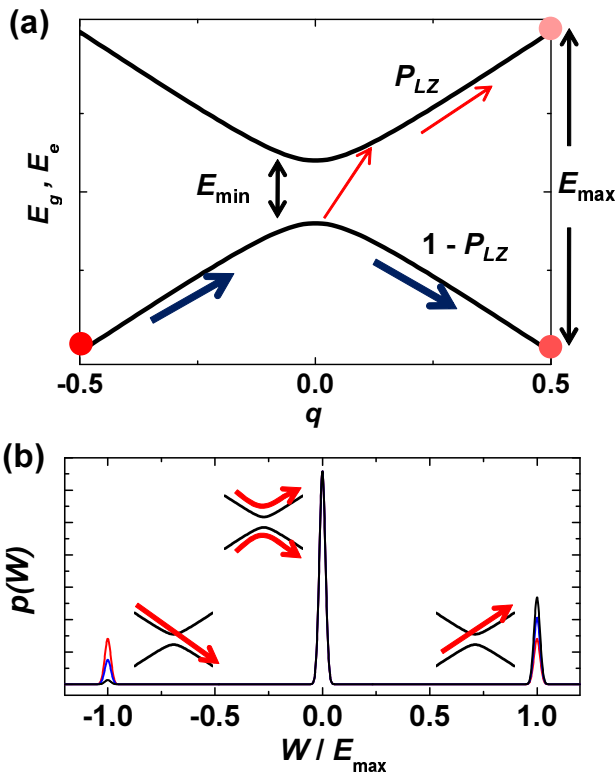}
    \caption{A two-level system with avoided crossing between the ground ($g$) and the excited ($e$) state. In (a) the control parameter $q$ is swept from left to right and the system is initially in the ground state and evolves either in the ground state (probability $1-P_{LZ}$) or makes a Landau-Zener transition (probability $P_{LZ}$). (b) Examples of the distribution of work in repeated sweeps under the conditions that the system starts in thermal equilibrium at various temperatures, and it is not coupled to the environment during the evolution. The distributions correspond to equation \eqref{w1} with artificial broadening of the peaks, and for $P_{LZ}=0.3$, and $\beta E_{\rm max} = 0.1,0.3$ and $1.0$ (red, blue and black curves, respectively). The peak at $W=-E_{\rm max}$ arises from $e$ to $g$ LZ transitions, the peak at $W=0$ is for adiabatic evolution, and that for $W=+E_{\rm max}$ arises from $g$ to $e$ transitions, as indicated by the diagrams within the (b) panel.}
    \label{fig1}
\end{figure}

Consider the basic setting for non-equilibrium fluctuation relations \cite{jarzynski97,crooks99}: initially the system is coupled to a bath and in thermal equilibrium with it. Thereafter it is driven by a control field in a non-equilibrium fashion. We are interested in the distribution of work performed on the system in repeated experiments with identical driving protocol among the realizations. In order to secure thermal equilibrium in the next realization, the interval between the repetitions is ideally infinitely long. We mainly consider the protocols where the drive signal returns to its initial position or to an equivalent one with equal free-energy at the end. Our main statement in this respect is then that, due to the energy conservation, the work $W$ done on the system under these conditions is equal to the heat $Q$ dissipated to the environment in the time interval from the beginning of the driving till the end of the equilibration period. With this premise we can avoid the formidable task of analyzing the work itself during the quantum evolution, and characterized by difficult-to-analyze time-orderings of the operators involved. The calorimetric principle can, at least in principle, be extended to more general protocols by bringing the system back to the initial position adiabatically at the end of the protocol.

Before going into the detailed analysis of the calorimetric method, let us first define the quantum system and present results in certain limiting cases for the work distribution, which can be obtained by known methods. We focus on a quantum two-level system, shown in Fig. \ref{fig1} and characterized by an avoided level crossing under the operation of the control parameter $q$. For simplicity, we consider a symmetric set-up, where the maximum energy spacing between the ground ($g$) and the excited ($e$) level is $E_{\rm max}$ in the beginning and the end of the drive, and $E_{\rm min}$ at the minimum in the middle. We normalize $q$ such that it obtains the values $\mp 1/2$ at the beginning and the end of the drive, respectively, and assumes a value $q=0$ at the avoided crossing. Suppose for the basic illustration that the system evolves in a unitary way from the initial state: this represents very weak coupling to the environment. In other words, the system is initially thermalized by coupling it to a bath with inverse temperature $\beta$, and decoupled from the bath for the duration of the driven evolution. This corresponds to a common situation for considering the fluctuation relations \cite{jarzynski97,campisi11}. The value of work in each realization is then determined by the internal energy stored in the system during the operation. There are two possibilities: the system makes a Landau-Zener (LZ) transition around $q=0$ between the two states with probability $P_{LZ}$, or it does not make the transition at probability $1-P_{LZ}$. If the system makes the transition, work $\pm E_{\rm max}$ is done on the system ($+$ for $g \rightarrow e$ transition, and $-$ for the $e \rightarrow g $ transition) by the end of the driving, otherwise the performed work vanishes.
The distribution of work $W$ can then be written as
\begin{eqnarray} \label{w1}
&& p(W)=(1-\rho_{gg}^0)P_{LZ}\,\delta(W+E_{\rm max})\nonumber\\&&+(1-P_{LZ})\,\delta(W)
+\rho_{gg}^0 P_{LZ}\,\delta(W-E_{\rm max}),
\end{eqnarray}
in agreement with the results in Refs. \cite{kurchan00,talkner07}. Here $\delta(E)$ refers to the Dirac delta function, and $\rho_{gg}^0=(1+e^{-\beta E_{\rm max}})^{-1}$ is the initial thermal population in the ground state. We may also state that the same distribution as for work applies for heat in the sense discussed above, since after an infinitely long time, for even weak coupling, the system has relaxed back to the thermal equilibrium state by exchanging energy with the environment. Examples of work distributions (with broadening) are shown in Fig. \ref{fig1} (b) at a few bath temperatures. As for the quantum systems evolving unitarily from a thermal state, the JE holds naturally for the two level system. Indeed we see immediately that
\begin{eqnarray} \label{w2}
\langle e^{-\beta W}\rangle = \int _{-\infty}^\infty dW p(W) e^{-\beta W} =1
\end{eqnarray}
for the distribution of equation \eqref{w1}. Here the brackets $\langle \cdot\rangle$ refer to averaging over the experimental realizations. Equations \eqref{w1} and \eqref{w2} simply confirm the earlier understanding of JE in a closed quantum system. The true challenge, which can be addressed by the here-proposed calorimetric detection, is, however, the measurement of work in an open quantum system coupled to the environment also during the driving period.

\begin{figure}
    \includegraphics[width=8.5cm]{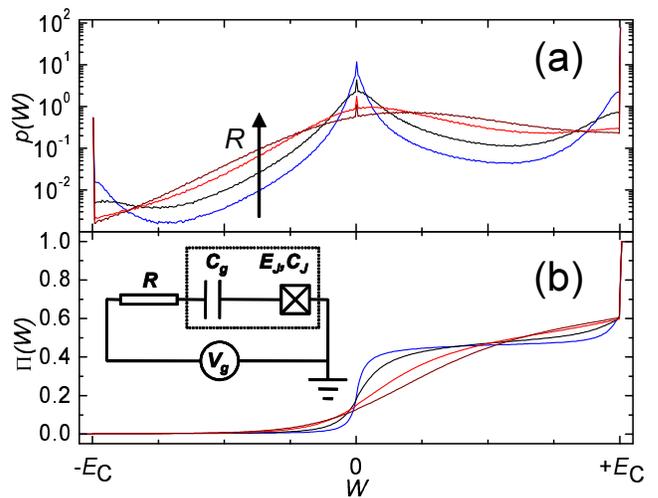}
    \caption{Numerical results of the distribution of work in a CPB. The circuit is shown in the inset of (b). In (a) we present the actual distribution $p(W)$. The parameters of the system are $E_J/k_B = 0.1$ K, $E_C/k_B = 1$ K, $\lambda =0.3$, bath temperature $k_B/\beta=0.2$ K, $\tau=1\cdot 10^{-9}$ s, and $R/R_Q = 1/3,1,10/3,10$. Each distribution is based on $9\cdot 10^7$ repetitions. In (b) we show the corresponding cumulative distributions $\Pi (W)=\int_{-\infty}^W p(W')dW'$ with the same colouring of the lines as in (a).
    }
    \label{results}
\end{figure}

Although our proposal and arguments are quite general, we specify next to a particular physical system, a Cooper-pair box (CPB), where concrete results about the distribution of work, and a description and analysis of the calorimetric measurement are immediately possible and can be implemented experimentally.
The CPB \cite{averin85,bouchiat98,nakamura99} consists of a superconducting island connected to a superconducting lead by a Josephson tunnel junction.
The system is described by the circuit scheme in the inset of Fig. \ref{results} and it is characterized by a voltage source $V_g$, coupling gate capacitance $C_g$, a Josephson junction with energy $E_J$ and capacitance $C_J$. We denote $C_\Sigma \equiv C_g + C_J$. Resistor $R$ forms the dissipative environment of the box.
In the regime $\epsilon \equiv E_J/(2E_C) \ll 1$, where $E_C=2 e^2/C_\Sigma$ is the charging energy of the box, we can treat the CPB as a two-level quantum system \cite{nakamura99}.
Denoting with $|0\rangle $ and $|1\rangle $ the states with zero and one excess Cooper-pairs on the island, respectively, the Hamiltonian of the system reads
\begin{equation}
 \hatH=  -E_C q (|1\rangle \langle 1| -|0\rangle \langle 0|)- \frac{E_J}{2} (|1\rangle \langle 0| +|0\rangle \langle 1|),
 \label{eq:CPB_Ham}
\end{equation}
where $q= C_g V_g/(2e)$ is the normalized gate voltage.
We assume a linear gate ramp $q(t)=-1/2+ t/\tau$ over a period $\tau=1/\dot q$ starting at $t=0$.
The energy gap separating the ground and the excited state of the system is given by $\Delta E = 2 E_C \sqrt{q^2+\epsilon^2}$. Thus the system passes a minimum energy gap $E_{\rm min} =E_J$ at $q=0$, see Fig. \ref{fig1} (a), where it makes a transition with the probability $P_{LZ}= \exp (-\frac{\pi \epsilon^2 E_C \tau}{\hbar})$ according to the standard LZ model \cite{landau32,zener32,shevchenko10}. The value of the energy gap at the ends of the trajectory is $E_{\rm max} =E_C\sqrt{1+(E_J/E_C)^2}\simeq E_C$.

We proceed further by exploiting the calorimetric principle to an open driven two-level system. The detailed dynamics of the CPB under the driven evolution with LZ transitions in the dissipative regime is beyond the focus in the present work. Instead, we consider the two level system in the incoherent regime, where energies of the order of the minimum gap $E_J$ or smaller are irrelevant. Then the system dynamics is governed by a master equation with rates between charge states $|0\rangle$ and $|1\rangle$ given by
\begin{equation} \label{rates}
\Gamma_\pm(\Delta E) = \frac{\pi}{2\hbar}E_J^2 {\mathcal P}(\pm \Delta E),
\end{equation}
where $\Delta E = 2E_C q$ is the energy dissipated (absorbed) in the $+$ ($-$) tunneling event and exchanged as a photon between the resistor and the system, and
\begin{equation} \label{pofe}
{\mathcal P}(E)=\frac{1}{2\pi \hbar} \int dt\exp({\mathcal J}(t)+\frac{i}{\hbar}Et)
\end{equation}
is the corresponding probability density of the system to emit energy $E$ to the environment \cite{ingold95}. The standard correlation function in equation (\ref{pofe}) is given by
\begin{eqnarray} \label{je}
&&{\mathcal J} (t) = \frac{4}{\pi}\int_0^\infty \frac{d\omega}{\omega}\frac{\Re {\rm e} Z_t(\omega)}{R_Q}\nonumber\\ &&\big\{\coth(\frac{1}{2}\beta\hbar\omega)[\cos(\omega t)-1]-i\sin(\omega t)\big\}.
\end{eqnarray}
Here, $R_Q \equiv \hbar/e^2$ and the relevant impedance of the CPB reads
\begin{eqnarray} \label{impedance}
\Re {\rm e} Z_t(\omega)=\frac{R_{\rm eff}}{1+(R_{\rm eff}C_{\rm eff}\omega)^2},
\end{eqnarray}
where $R_{\rm eff}=\lambda^2 R$, $C_{\rm eff}=(\lambda^{-1}-1)C_\Sigma$, and $\lambda \equiv C_g/C_\Sigma$ is the coupling parameter.
To obtain the distribution of work, our task then is to keep track on the net energy deposited to the resistor in each ramp.

We have analyzed the distribution of work numerically by stochastic simulations tracing the paths, assuming rates of equation \eqref{rates}. Results of such simulations are shown in Fig. \ref{results}. For all these distributions JE is numerically valid within $0.1$ $\%$, and it should naturally be valid identically since ${\mathcal P}(E)$, and thus the $\pm$ rates in equation \eqref{rates} obey detailed balance \cite{crooks98}. These simulations correspond to the pumping trajectory described above, with parameters given in the Fig. \ref{results} caption. The distributions have several features that can be explained by simple arguments as follows. First of all, the peaks in (a) and the steps in (b) at $W=\pm E_C$, and at $0$ correspond to the peaks arising from excitation/relaxation, and from the adiabatic dynamics of the unitary evolution in equation \eqref{w1}, respectively, and they get gradually weaker towards faster relaxation (i.e., towards increasing $R$). In general, the mean of the work is positive, i.e. the evolution is dissipative. Furthermore, we have checked numerically that the Crooks equality \cite{crooks99} is valid in the form $p(-W)=e^{-\beta W}p(W)$ for these distributions.

\begin{figure}
    \includegraphics[width=8.5cm]{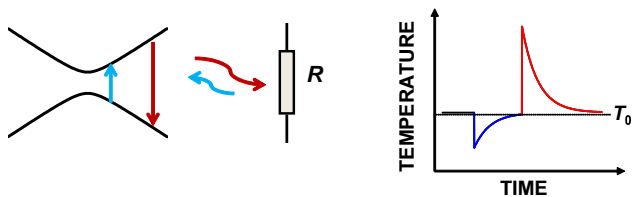}
    \caption{Schematic presentation of the calorimetric measurement. The resistor in the middle emits (absorbs) heat to (from) the two-level system on the left upon excitation (relaxation) shown by blue (red). The corresponding heat pulses, i.e. temperature of the electrons in the resistor vs. time, are shown on the right with the same colour conventions. We assume that the electrons are in the so-called hot-electron regime, where the internal (electron-electron) relaxation is faster than other time scales in the problem.
    }
    \label{calorimetry}
\end{figure}

Next we discuss the calorimetric measurement of the energy deposited on the resistor. The electronic system of the resistor forms the environment that is then coupled to a super-bath formed, e.g. by phonons to which these electrons couple by standard electron-phonon coupling. This leads to the following sequence once an energy quantum $\Delta E$ is emitted/absorbed by the two-level system due to $\mp$ transitions, see Fig. \ref{calorimetry}.  We assume the standard hot-electron regime \cite{richards94,giazotto06}. (i) The temperature $T$ of the resistor (electrons) changes abruptly by $\Delta T = \pm \Delta E /\mathcal{C}$ at the moment of the photon exchange; here $\mathcal{C}$ is the heat capacity of the electron gas in the resistor. (ii) The electronic temperature of the resistor starts to relax back towards the equilibrium $T_0 \equiv 1/(k_B\beta)$ at the rate governed by the thermal conductance $G$ that is determined, e.g. by electron-phonon coupling. (iii) New energy pulses may occur during the evolution. In this evolution the deposited energy (heat) to the bath can be obtained generally from $\dot Q = \mathcal{C}\dot T + G\Delta T$, where $\Delta T = T - T_0$. Note that the pulses can have either $+$ or $-$ sign, i.e., the resistor may cool as well. In the considered example, $\dot Q(t)=\sum_i \Delta E_i \delta(t-t_i)$, where $\Delta E_i$ is the magnitude of heat released in the $i$th absorption/emission event, and $t_i$ are the time instants of these events. Note that unlike in the analysis leading to equations (\ref{w1}) and (\ref{w2}), we allow here different values for $\Delta E_i$ for transitions occuring along the driven trajectory. Integrating the equation for $\dot Q$ for the situation where the system is in thermal equilibrium in the beginning and at the end, i.e. $T(0)=T(\infty)=T_0$, and assuming small temperature rise in the pulses, we have for the heat in each realization
\begin{equation} \label{cal}
Q=G\int_0^\infty \Delta T(t) dt.
\end{equation}
To make full use of equation (\ref{cal}), the following practical issues need to be taken care of \cite{giazotto06}. The thermometer is first calibrated against the bath temperature under the equilibrium conditions (without applying the driving field), or one uses a primary thermometer that serves without calibration. The only remaining calibration is then the determination of $G$: this can be obtained by applying a known constant power $P$ to the calorimeter, and measuring the temperature rise $\delta T$ of the electrons in the thermometer, since $G=P/\delta T$.

Before turning to the detailed practical realization and numbers, one further consideration is in order. Namely, biasing the gate capacitor in the ramp that we consider leads to a current that dissipates Joule power $\Delta E_d$ in the resistor $R$. This dissipation can be estimated by elementary circuit analysis with the result $\Delta E_d =e^2R/\tau$. For typical values of $\tau$ and $R$, for instance those in Fig. \ref{results}, this energy is at most of order $0.01E_C$. Since $E_C$ is the scale of work performed, the Joule heating produces only a small error in the calorimetric measurement, even if no precautions are taken.

Finally we analyze the feasibility of measuring calorimetrically the distribution of quantum work in a superconducting two-level system considered. The most straightforward realization would involve a small lithographic metal resistor, $\sim (0.1$ $\mu$m$)^3$, on the chip \cite{timofeev09}. The electronic heat capacity at $T_0=100$ mK temperature is $\mathcal{C} \sim 10^{-20}$ JK$^{-1}$, yielding a jump in temperature of $E_C/\mathcal{C} \sim 3$ mK upon absorbing the relaxation heat at the end of the ramp, assuming a realistic $E_C/k_B = 3$ K. This is a sizable change of temperature (3\%) which can be detected over a relatively long time of order $10^{-4}$ s that is determined by the equilibration of the resistor back to its initial temperature via electron-phonon coupling. The large sensitivity of the calorimeter is naturally due to the standard $T_0/T_F\sim 10^{-6}$ suppression of the heat capacity of the electron gas with Fermi temperature $T_F\sim 10^5$ K. If needed, longer relaxation times can be obtained by, for instance, suspending the resistor on a silicon nitride membrane \cite{vercruyssen11}, which would make the detection even easier. A tunnel probe, in form of a superconducting lead connected to the resistor, i.e. a SIN junction, measured by RF reflection or transmission techniques \cite{cleland2003}, is fully compatible with the requirements of the presented calorimetric detection of heat.


{\bf Acknowledgements} We thank J. Ankerhold, S. Gasparinetti, D. Golubev, F. Hekking, M. M\"ott\"onen, and J. Santos for discussions. The work has been supported
partially by LTQ (project no. 250280) CoE grant and the
European Community's Seventh Framework Programme
under Grant Agreement No. 238345 (GEOMDISS).

\end{document}